\documentclass[12pt,a4paper]{article}

\usepackage[left=2cm,right=2cm,top=2cm,bottom=2cm]{geometry}

\usepackage{graphicx}  
\usepackage{dcolumn}   
\usepackage{bm}        
\usepackage{amssymb}   
\usepackage{amsmath}

\usepackage{cite}

\usepackage{indentfirst}

\usepackage{authblk}

\usepackage{cmap}

\usepackage{float}
\usepackage[caption = false]{subfig}
\newcommand{\guu}{g_{\uparrow \uparrow}}
\newcommand{\gdd}{g_{\downarrow \downarrow}}
\newcommand{\gud}{g_{\uparrow \downarrow}}
\newcommand{\gdu}{g_{\downarrow \uparrow}}
\newcommand{\fud}{f_{\uparrow \downarrow}}
\newcommand{\fdu}{f_{\downarrow \uparrow}}
\newcommand{\fuu}{f_{\uparrow \uparrow}}
\newcommand{\fdd}{f_{\downarrow \downarrow}}

\begin{document}

\title{Density of states in the heterostructure ferromagnetic insulator--superconductor--ferromagnetic insulator}

\author[1]{D.V. Seleznev}
\author[1]{S.S. Seidov}
\author[1]{N.G. Pugach}
\author[1]{D.G. Bezymiannykh}
\author[2]{S.I. Mukhin}
\author[1]{B.G. L'vov}

\affil[1]{HSE University, Moscow, Russia}
\affil[2]{NUST MISIS, Moscow, Russia}

\date{}

\maketitle

\begin{abstract}

We consider a spin valve composed of a superconducting film (S) between two ferromagnetic insulators (FI) on two sides. In the dirty limit the superconductor is described by Usadel equations. Appropriate boundary conditions were chosen for two S--FI interfaces, which are described via the interface parameter spin mixing angle. By numerically solving the Usadel equations, the density of states (DOS) at different spin mixing angles were obtained. It was shown previously that critical temperature of such FI--S--FI structure depends on the mutual alignment of the FI layers magnetization. We follow the evolution of DOS at change of misalignment of ferromagnets magnetization and probe the zero bias peak creation. The DOS characteristic features may give a fruitful information about triplet superconducting components creation and interplay inside the S layer.\\

\end{abstract}

\section{Introduction}
Spin valves are heterostructures of magnetic and superconductor layers which can be used to control the passage of the supercurrent due to the proximity effect. These devices lie in the foundation of superconducting spintronics \cite{linder_superconducting_2015, leksin_peculiarities_2013, leksin_boosting_2016, kamashev_superconducting_2019, gordeeva_record_2020, heim_ferromagnetic_2013} which aims to develop alternative microelectronics based on superconductors. The spin valve may be composed by different arrangement of magnetic and superconducting layers and by choosing different types of magnets. In particular, one might place the superconductor between two ferromagnets (FSF) or place the superconductor next to two ferromagnets, separated by a non--magnetic material (SFF) \cite{birge_spin-triplet_2018, eschrig_theory_2018, golovchanskiy_magnetization_2020, azizi_crossed_2022, he_spin-related_2023, melnikov_superconducting_2022, karelina_magnetic_2024}. The ferromagnets might be either conducting or insulating, weak or strong. Quantitatively the strength of the magnet is determined by the energy splitting of spin subbands compared to the Fermi energy. Alternatively one might use a single ferromagnet with a controlled non--collinear magnetic structure \cite{gusev_magnonic_2021}.

The proximity effect at the interface between the superconductor and the ferromagnet leads to conversion of the singlet correlations from the bulk of the superconductor into triplet ones \cite{buzdin_proximity_2005, bergeret_odd_2005, de_gennes_boundary_1964, eschrig_general_2015}. At the same time oscillations of the paired correlation function occur in the ferromagnet itself. Contact with a single ferromagnet generates only the triplet correlations with zero spin projection. In order to additionally create triplet correlations with spin projections $\pm 1$ non--collinear magnetization in the structure is required. Said correlations provide a long--range Josephson effect in ferromagnetic Josephson junctions \cite{vedyayev_critical_2006, klenov_examination_2008}, the triplet effect of the spin valve \cite{flokstra_controlled_2015} and assumed to be cause of the ``giant’’ spin valve effect \cite{aarts_colossal_2015}. However the effect might have a different mechanism in the ferromagnetic insulator--superconductor--ferromagnetic insulator heterostructure \cite{blamire_superconducting_2017}.

In the present work we consider an FI--S--FI spin valve (FI --- ferromagnetic insulator). The interplay of two FI layers allows creation of all possible triplet correlations in the S layer \cite{tanaka_2007, linder_2024}. As mentioned in the previous paragraph, when the magnetization axes of the ferromagnets are parallel, only the triplet correlation with zero spin projection is generated. The rest are generated when the magnetization is non--collinear, i.e. the ferromagnets are misaligned. The order parameter is suppressed at the interfaces due to proximity effect. The critical temperature is the lowest at the parallel orientation of the magnetization directions of the FI layers and is maximal when the magnetizations of the FI layers are antiparallel \cite{blamire_superconducting_2017}. The thickness of the S layer should be of the order of the coherence length --- this allows the interplay of triplets penetrating from both sides and is still thick enough for the superconductivity not being completely suppressed.

It is convenient to study such systems in the quasiclassical approximation \cite{larkin_quasiclassical_1969, eilenberger_transformation_1968}. In the dirty limit, the matrix Green's function is described by Usadel equations \cite{usadel_generalized_1970}. The dirty limit conditions, i.e. the superconducting coherence length is much larger than the electron mean free path, usually correspond to experimental situation of superconducting heterostructures made by sputtering. In order to describe a spinactive interface one should impose boundary conditions on the Usadel equations, which turn out to be highly nontrivial \cite{zaitsev_quasiclassical_1984, kurpianov_influence_1988, linder_quasiclassical_2022, eschrig_general_2015}. If the interface is not conducting, i.e. the magnetic material is an insulator, the exact expression for the boundary condition is known \cite{ouassou_triplet_2017}. However, exact analytical solution of the Usadel equations is complicated. One could either employ numerical methods or solve the equations in linear approximation at the temperature close to the temperature of the superconducting phase transition, when the anomalous components of the Green's function are small. Change of the critical temperature in the FSF superconducting spin valve was calculated theoretically \cite{fominov_triplet_2003}, however the detailed analysis of the triplet superconducting correlations via study of the density of states (DOS) in the superconductors was not carried on neither for weak ferromagnets nor for strong ones.

In our previous papers \cite{yagovtsev_magnetization_2020, yagovtsev_magnetization_2021, seleznev_reverse_2023, seleznyov_ferromagnetic_2023, yagovtsev_inverse_2021} we have considered a superconductor in contact with a single ferromagnet. In the present manuscript we investigate the heterostructure ferromagnetic insulator--superconductor--ferromagnetic insulator. The Usadel equations, describing it, are solved numerically for an arbitrary angle of ferromagnets misalignment. The spatial distributions of anomalous components of the Green's functions in the superconductor were obtained. From these distributions we calculated the DOS in the superconductor as a function of the misalignment angle of the ferromagnets, spin mixing angle and thickness of the superconductor, as it provides great insight in the physics of the superconductor \cite{khaydukov_magnetic_2011, salikhov_spin_2009, gupta_anomalous_2004, halterman_half-metallic_2016, alidoust_half-metallic_2018, banerjee_evidence_2014}. The results witness the existence of triplet correlations in the superconductor, which are generated by the ferromagnetic insulators. Although this is what one expects, the novelty is in the following: usually the triplet correlations are investigated by studying their influence on the critical temperature, which is an integral characteristic. The approach, used in the present paper, has a microscopic nature and is also applicable at arbitrary temperatures.

\section{Usadel equations and boundary conditions}
A superconductor between two ferromagnet insulators from both sides is described by the matrix Green's function
\begin{equation}
\hat g = \begin{pmatrix}
\guu & \gud & \fuu & \fud\\
\gdu & \gdd & \fdu & \fdd\\
-\fuu & -\fud & -\guu & -\gud\\
-\fdu & -\fdd & -\gdu & -\gdd
\end{pmatrix}.
\end{equation}
Ferromagnetic insulators from left and right have spin mixing angles $\phi_L$, $\varphi_R$ and magnetic moments, oriented in directions $\mathbf{m}_L$, $\mathbf{m}_R$ respectively.

In the diffusive limit the spatial distribution of the Green's function in the superconductor is described by the Usadel equation
\begin{equation}\label{eq:Usadel}
\begin{aligned}
&i D \partial_z (\hat g \partial_z \hat g) = [\varepsilon \tau_3 \otimes \sigma_0 + \hat \Delta, \hat g]\\
&\hat \Delta = \operatorname{antidiag}(+\Delta, -\Delta, +\Delta^*, -\Delta^*).
\end{aligned}
\end{equation}
Here, $D$ is the electron diffusion constant in the superconductor, $\varepsilon$ is the energy of the quasiparticles, $\sigma$ and $\tau$ are the Pauli matrices in spin and Nambu spaces respectively, $\Delta$ is the superconducting gap.

Usadel equations should be complemented by appropriate boundary conditions. In general, they take the form
\begin{equation}
G_L L_L (\hat g_L \partial_z \hat g_L) = -G_R L_R (\hat g_R \partial_z \hat g_R) = I.
\end{equation}
Indices $L$ and $R$ correspond to the left and right borders of the superconductors, matrix $I$ in the right hand side is called the matrix current, $G_{L/R}$ and $L_{L/R}$ are the conductance and length of the materials respectively. The interface between the superconductor and the ferromagnetic is a spinactive non--conducting one, which is describe by matrix current \cite{ouassou_triplet_2017}
\begin{equation}\label{eq:I}
\begin{aligned}
I &= - N G_Q \left[1 - \frac{i}{4} \sin(\phi) \hat a + \frac{1}{2} \sin^2 \left(\frac{\phi}{2} \right) \hat a \hat m \right]^{-1} \times\\
&\times \left[-i \sin(\phi) \hat g \hat a + \sin^2 \left(\frac{\phi}{2} \right) [\hat m, \hat a] \right] \times\\
&\times \left[1 - \frac{i}{4} \sin(\phi) \hat a + \frac{1}{2} \sin^2 \left(\frac{\phi}{2} \right) \hat m  \hat a \right]^{-1}.
\end{aligned}
\end{equation}
Here, $G_Q = e^2/\pi$ is the conductance quantum, $N$ is the number of the scattering channels at the interface, $\phi$ is the spin mixing angle of the ferromagnetic insulator, $\hat a = \hat g \hat m \hat g - \hat m$ and for the magnetic moment orientation, $\mathbf{m} = (m_x, m_y, m_z)$ of the ferromagnetic insulator, the matrix $\hat m$ is
\begin{equation}
\hat m = \begin{pmatrix}
1 & 0\\
0 & 0
\end{pmatrix} \otimes
\begin{pmatrix}
m_z & m_x - i m_y\\
m_x + i m_y & -m_z
\end{pmatrix} +
\begin{pmatrix}
0 & 0\\
0 & 1
\end{pmatrix} \otimes
\begin{pmatrix}
m_z & m_x + i m_y\\
m_x - i m_y & -m_z
\end{pmatrix}.
\end{equation}

\section{Numerical results and discussion}
The numerical procedure consists of solving Usadel equations (\ref{eq:Usadel}) with boundary conditions (\ref{eq:I}) and self--consistency equation 
\begin{equation}
\Delta(z) = \frac{N_0 \lambda}{2} \int \limits_{0}^{\Delta_0 \cosh(1/N_0 \lambda)}\Big\{f_s(z, \varepsilon) - f_s(z, -\varepsilon)  \Big\} \tanh \left( \frac{\varepsilon}{2T} \right) d \varepsilon.
\label{selfcons}
\end{equation}
Here $N_0$ is the density of states at the Fermi level, $\lambda$ is the BCS coupling constant, $T$ is the temperature, $f_s = (\fud - \fdu)/2$ is the singlet component of the anomalous Green's function. This allows to calculate the Green's function $\hat g(z)$ and the density of states
\begin{equation}
N(z, \varepsilon) = \operatorname{Re} \left( \frac{\guu + \gdd}{2} \right)
\end{equation}
as a function of the energy of the quasiparticles, angle of ferromagnets magnetic moments misalignment $\theta$ and the spin mixing angle $\phi$. Further all the values of the DOS will be taken at the center of the superconductor in order to achieve equal contribution from the left and right ferromagnets. Next we plot the DOS and study the plots, the presented results are at the temperature $T = 0.08 T_c$ and the thickness of the superconductor $L = 1.2\xi$, where $\xi$ is the cooper pair coherence length.

\begin{figure}[h!!]
\subfloat[a) $\theta = 0$]{\includegraphics[width = 0.49\textwidth]{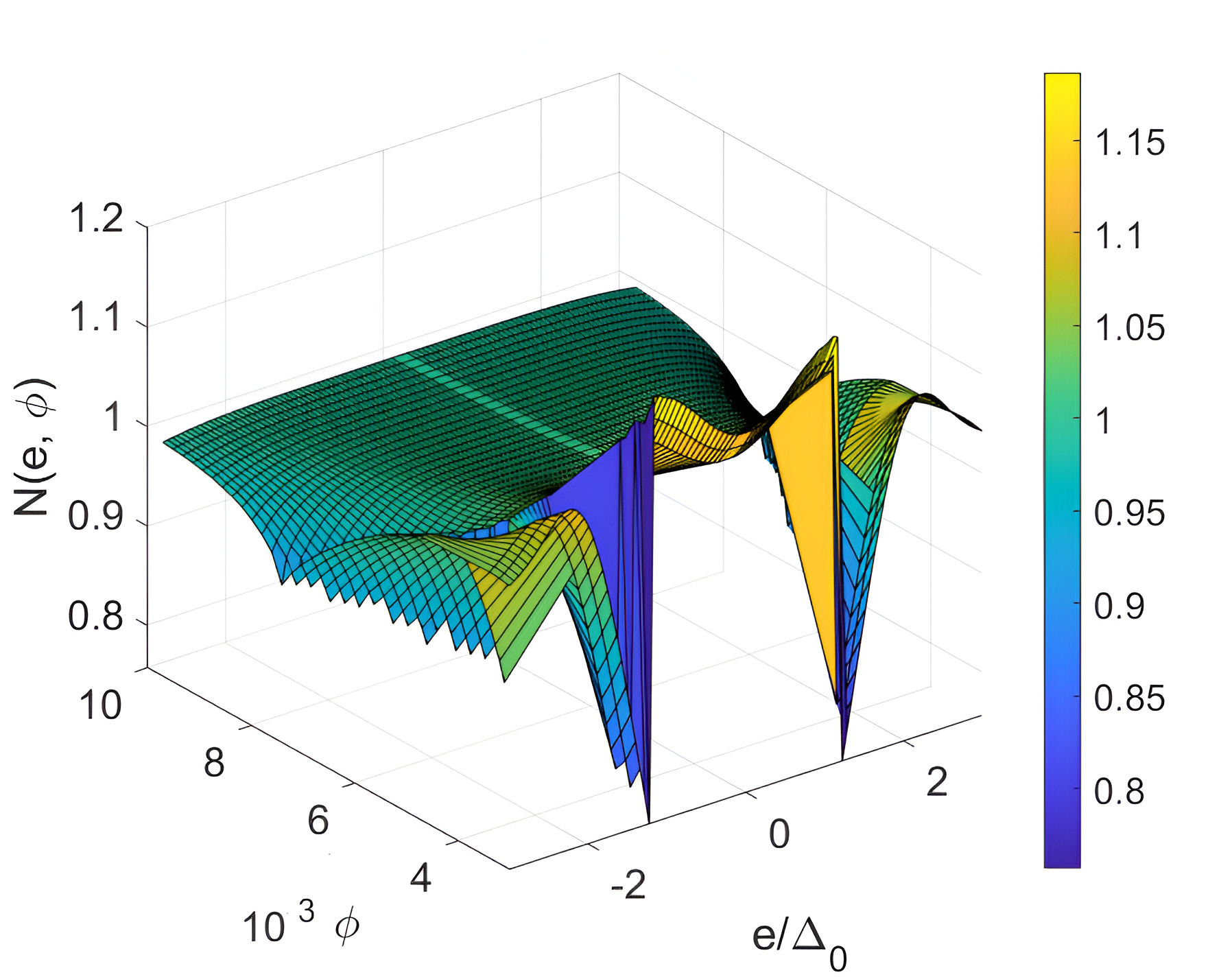}}
\subfloat[b) $\theta = \pi/2$]{\includegraphics[width = 0.49\textwidth]{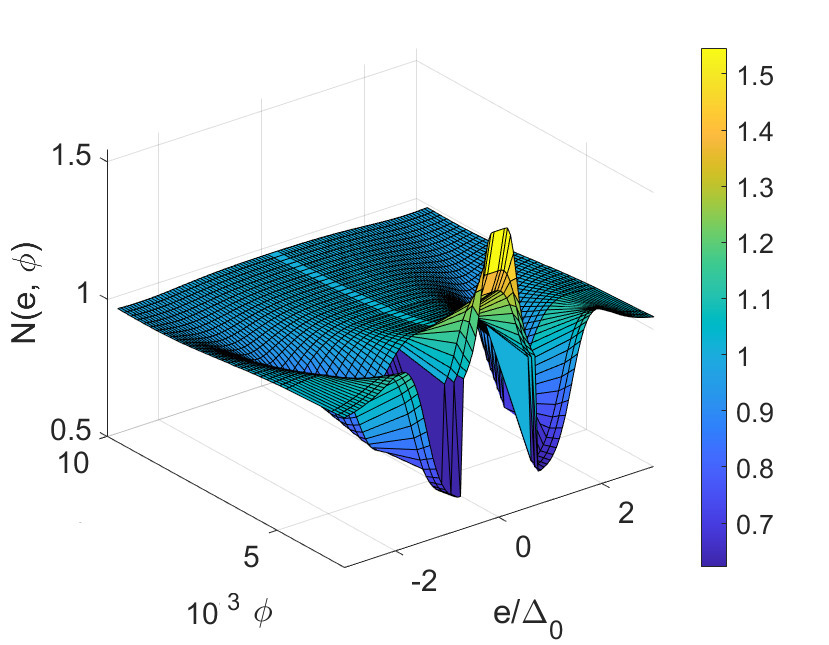}}\\
\center\subfloat[c) $\theta = \pi$]{\includegraphics[width = 0.49\textwidth]{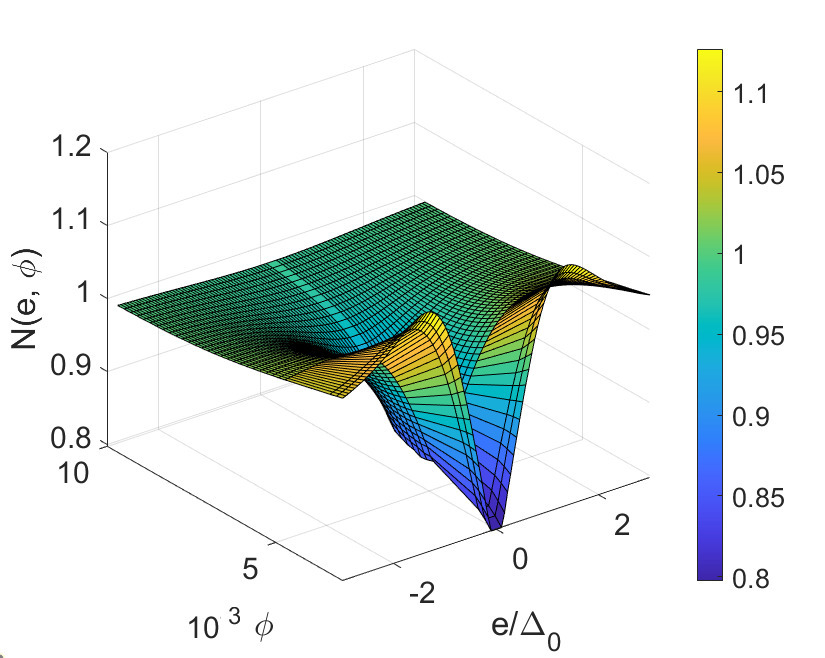}}
\caption{Dependence of the density of states on the energy of the quasiparticles as function of spin mixing angle $\phi$ for angle of ferromagnets magnetic moments misalignment $\theta = 0$ (a), $\theta = \pi/2$ (b) and $\theta = \pi$ (c).}
\label{fig:DoS_3d}
\end{figure}

In fig. \ref{fig:DoS_3d} the dependence of density of states on the energy of quasiparticles and spin mixing angle are plotted for $\theta = 0$, $\theta = \pi/2$ and $\theta = \pi$.
First of all, one can observe that the superconducting gap closes at high spin mixing angles $\phi$. In this case singlet correlations are converted to the triplet ones and the singlet s--wave superconductivity becomes similar to the ferromagnetic superconductor type.

At $\theta = 0$ in fig. \ref{fig:DoS_3d}a) the BCS gap, typical for s--wave superconductors, is closed, which is due to singlet Cooper pairs converting to triplet ones in the superconductor, induced by spin mixing at the interface. This process suppresses the superconducting gap. Triplet correlations also lead to spreading of the zero bias peak (ZBP) along the whole gap.

In the case of perpendicular orientation of the magnetization directions at $\theta = \pi/2$, which is presented in fig. \ref{fig:DoS_3d}b), one can observe a sharp ZBP, which the a sign of existence of the triplet correlations in the superconductor. In this configuration triplet correlations with all three possible spin projections $-1$, $0$ and $1$ are generated. We will study the dependence of the ZBP on the angle $\theta$ further in the manuscript.

Finally, in the case of the antiparallel orientation of the magnetization directions at $\theta = \pi$, which is plotted in fig. \ref{fig:DoS_3d}c), the superconducting gap is alike the smoothed BCS one. Contributions from the triplet proximity effect from the two boundaries partially compensate each other, the superconductivity is suppressed the least and the critical temperature of the superconductor is maximal \cite{blamire_superconducting_2017}.

\begin{figure}[h!!]
\center\includegraphics[width = 0.49\textwidth]{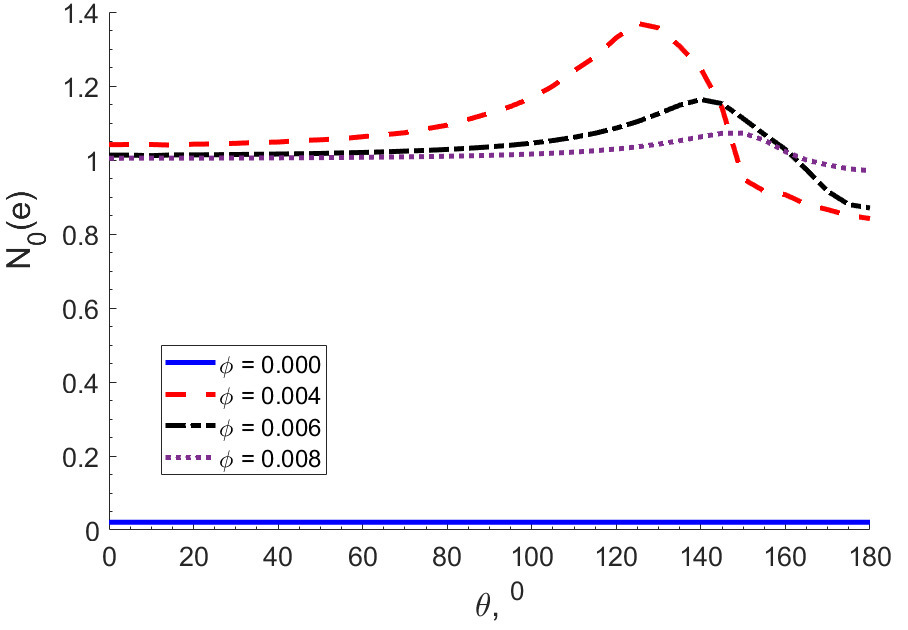}
\caption{Dependence of the density of states at the Fermi level on the angle of ferromagnets magnetic moments misalignment $\theta$ for different spin mixing angles $\phi$.}
\label{fig:DoS_2d}
\end{figure}

In order to study the density of states as function of angles $\phi$ and $\theta$, we have plotted its value at the Fermi level as function of $\theta$ for different values of $\phi$, fig. \ref{fig:DoS_2d}.
As expected, the density of states at the Fermi level is zero for zero spin mixing angle. Next, even at small non--zero $\phi$ the DOS at the Fermi level spikes rapidly diverging from zero and reaches it maximum at some angle $0 < \phi < 0.006$. For larger spin mixing the DOS reduces reaching unity, as discussed earlier and can also be seen in fig. \ref{fig:DoS_3d}. The dependence on $\theta$ is also non--monotonic, having a maximum, which is the zero bias peek. The later is placed not at $\theta = \pi/2$, but is shifted to the right. This is due to interplay between singlet and triplet correlations in the superconductor \cite{seleznyov_ferromagnetic_2023}. In particular, the contribution to the DOS of the correlations with spin projections $\pm 1$ has a maximum at $\theta = \pi/2$, which becomes shifted by the contribution from correlations with spin projections $0$.

\section{Conclusions}
This paper continues our previous studies of superconducting heterostructures with magnetic materials. In those, we have considered a superconductor with a magnetic material at one of its boundaries, whereas the second one remained free. Now, we have considered a superconductor with two ferromagnetic insulators at both sides. The interplay of triplet correlations, created by ferromagnets, leads to nontrivial effects in the superconductor, depending on the angle of misalignment of magnetic moments. Based on the Usadel equations, the density of states in the superconductor was found. The ferromagnets were represented as nonconducting spin-active interfaces with corresponding boundary conditions for the Usadel equations.

Features of the DOS, such as suppressed BCS gap and the ZBP, demonstrate generation of triplet correlations in the superconductor due to inverse proximity effect at the boundary. The properties of the DOS and consequently the critical temperature of the superconductor can be controlled changing the magnetic configuration of the heterostructure. This is the mechanism, based on which one could design a superconducting spin valve. Namely, the superconductor can be brought from superconducting to the normal state and vice versa, closing and opening transition of the superconducting current, by changing the magnetic configuration.

Dependence of the ZBP at the Fermi level on the angle of magnetizations misalignment and spin mixing angle was studied separately. As one expects, there is no triplet proximity effect at zero spin mixing angle and, if the spin mixing angle is large enough, the superconductivity is completely suppressed.

Theoretical study of the DOS of the superconducting layer in a FI--S--FI spin valve provides knowledge about the nature of triplet correlations. The latter, generated due to interaction with the FI layers, are responsible for suppression of the superconducting gap and the spin valve effect. Thus understanding of their physics allows to optimize the design of spin valve devices.

\section{Acknowledgements}
The publication was prepared within the framework of the Academic Fund Program at HSE University \#24-00-038 ``Quasiclassical dynamics of quantum systems: chaotic and spatially nonhomogeneous like heterostructures superconductor--magnetic''.

\bibliographystyle{ieeetr}

\end{document}